\documentstyle[12pt,aaspp4,flushrt]{article}     

\begin{document}

\title{
A SUB-MILLIMETER-WAVE ``FLARE'' FROM GG TAU?
}

\author{Gerald H. Moriarty-Schieven}
\affil{Joint Astronomy Centre, 660 N. A'ohoku Pl., Hilo, HI  96720 
\& Dominion Radio Astrophysical Observatory, 
National Research Council of Canada (gms@jach.hawaii.edu)}

\and

\author{Harold M. Butner}
\affil{Department of Terrestrial Magnetism, Carnegie Institution of 
Washington, 5241 Broad Branch Rd., NW, Washington, DC \,\,  20015 
(butner@dtm.ciw.edu)}

\begin{abstract}

We have monitored the millimeter and submillimeter emission from the 
young stellar object GG Tau, a T Tauri binary system surrounded by a
massive circumbinary disk.  We find that between 1992 and 1994,
the flux has increased significantly at 800, 1100, and 1300 ${\mu}m$, resulting
in a steepening of the observed spectral energy distribution
at those wavelengths.
Such an increase appears consistent with a modest increase in
disk luminosity (a factor of two).  The increase in the effective
disk temperature might arise from a slight change in the disk
heating processes.  Alternatively, the flux increase may reflect
a sudden change in the underlying dust optical properties.

\end{abstract}

\keywords{radio continuum: stars -- 
stars: flare -- stars: formation -- stars: individual: GG Tauri}

\pagebreak

\section{Introduction}

GG Tau is a young stellar system within the Taurus molecular cloud complex
which is located at a distance of $\sim$140pc (Elias 1978).  It consists
of a pair of close binary stars (Leinert et al. 1991) separated by $\sim$10''
($\sim$1500 AU).  The main component of this binary pair consists of a 
K7-M0 star (Strom et al. 1989; Hartigan et al. 1994) and an M4 star 
(Roddier et al. 1996) separated by 0.255'' ($\sim$38 AU), and has collectively
been classified as an emission-line (``classical'') T Tauri star (Herbig
\& Bell 1988).  Associated with this system, which we will refer to simply as
GG Tau, is a large, massive
circumstellar disk, which has recently been
imaged using millimeterwave interferometers by a number of groups (Simon
\& Guilloteau 1992;  Kawabe et al 1993;  Koerner, Sargent \& Beckwith 1993;
Dutrey, Guilloteau \& Simon 1994).  They found a well-resolved, extended 
feature, whose morphology and kinematics are consistent with a rotating
{\em circumbinary} disk (Dutrey, Guilloteau \& Simon 1994; Koerner,
Sargent \& Beckwith 1993).  Dutrey et al. have also found that the disk, of
outer radius $\sim$800 AU, appears to have a central hole or cavity of 
radius $\sim$180 AU.  Roddier et al. (1996) have recently presented 
near-infrared images acquired using adaptive-optics techniques, which clearly
show a partially evacuated cavity of radius $\sim$200 AU in the disk.  They
also find some infrared excess from both of the stars inside the cavity, 
perhaps from warm inner disks surrounding each star.

The circumstellar/circumbinary disk radiates very strongly at millimeter
and sub-millimeter wavelengths (Beckwith et al 1990;  Beckwith \& Sargent 1991;
Moriarty-Schieven et al 1994).  Indeed, the published 800 ${\mu}m$ flux
densities are nearly 25\% of the {\em IRAS} 100 ${\mu}m$ flux density of 5.2
Jy.  We present here evidence that the sub-mm flux density of GG Tau has
varied.  We interpret our observations as a sudden 
brightening or ``flare'' of the disk.
We will present a set of simple models that can match the observations,
and discuss their implications.

\section{Observations}
\begin{planotable}{ccccc}
\tablecaption{GG Tau Flux Densities}
\label{flux-tab}
\tablehead{
 \colhead{${\lambda}$  $({\mu}m)$}	&	\colhead{Date of Obs.}	&
\colhead{Beam FWHM}	&	\colhead{$S_{\nu}$ (Jy)}	&	
\colhead{References}
}
\startdata
12    &       1983   &   25''    &   1.27 $\pm$ 0.13   & IRAS PSC \nl \nl

25    &       1983   &   25''    &   1.65 $\pm$ 0.15   & IRAS PSC \nl \nl

60    &       1983   &   60''    &   3.03 $\pm$ 0.27   & IRAS PSC \nl \nl

100   &       1983   &   100''   &   5.16 $\pm$ 0.57   & IRAS PSC \nl \nl  
450   &   Jan 1994   &   17.5"   &   4.54 $\pm$ 0.35   &   MSB   \nl  
      &   Dec 1994   &   17.5"   &   4.16 $\pm$ 0.23   &   MSB   \nl  \nl

800   &   29 Nov, 3 Dec 1989, 7 Dec 1990   &   17"   &   1.25 $\pm$ 0.08   &
   BS   \nl
      &   Sept 1991   &   16.5"   &   1.11 $\pm$ 0.12   &   MSWKT   \nl
      &   Jan 1994   &   16.5"   &   1.71 $\pm$ 0.04   &   MSB   \nl
      &   Sept 1994  &   16.5"   &   1.59 $\pm$ 0.06   &   MSB   \nl   
      &   Dec 1994   &   16.5"   &   1.65 $\pm$ 0.08   &   MSB   \nl   \nl

1100  &   2 Dec 1989 &   22"     &   0.800 $\pm$ 0.051 &   BS    \nl
      &   Sept 1991   &   18.5"   &   0.74 $\pm$ 0.12   &   MSWKT \nl
      &   Jan 1994   &   18.5"   &   1.07 $\pm$ 0.03   &   MSB   \nl
      &   Sept 1994  &   18.5"   &   0.83 $\pm$ 0.03   &   MSB   \nl   
      &   Dec 1994   &   18.5"   &   0.85 $\pm$ 0.03   &   MSB   \nl   \nl

1300  &   29, 30 April 1988 & 11" &  0.593 $\pm$ 0.053 &   BSCG  \nl
      &   Jan 1994   &   19"     &   0.69 $\pm$ 0.03   &   MSB   \nl
      &   Sept 1994  &   19"     &   0.63 $\pm$ 0.02   &   MSB   \nl   
      &   Dec 1994   &   19"     &   0.63 $\pm$ 0.03   &   MSB   \nl   \nl

2000  &   Jan 1994   &   28"     &   0.32 $\pm$ 0.06   &   MSB   \nl

\enddata
\tablerefs
{IRAS PSC = {\em IRAS} Point Source Catalog v. 2.1;
BS = Beckwith \& Sargent (1991);  BSCG = Beckwith et al (1990);  
MSB = this work;  MSWKT = Moriarty-Schieven et al (1994)}

\end{planotable}
The data to be discussed in this report are presented in Table 1, and include
new observations as well as observations of Beckwith et al (1990; hereafter
BSCG), Beckwith \& Sargent (1991; hereafter BS), and Moriarty-Schieven
et al (1994; hereafter MSWKT).
The new observations presented here were obtained at the James Clerk 
Maxwell Telescope (JCMT)\footnote{The JCMT 
is operated by the Royal Observatory Edinburgh on
behalf of the Particle Physics and Astronomy Research Council
of the United Kingdom, the Netherlands
Organization for Scientific Research, and the National Research Council of
Canada.} located near the summit of Mauna Kea, in January, September, and
December
1994.  The photometry was obtained using the UKT14 bolometer equipped with
filters passing
effective wavelengths of 444, 790, 1090, 
1260 and 1920 ${\mu}m$, using a 65mm aperture (corresponding to 17.5", 
16.5", 18.5", 19" and 28" (FWHM) 
respectively on the sky), chopping in azimuth
by 60" at a frequency of 7.8 Hz.

\begin{planotable}{cccc}
\tablecaption{Flux Densities of Secondary Calibrators}
\label{crl618-tab}
\tablehead{
 \colhead{Source}	&	\colhead{$\lambda$ (${\mu}m$)}	
 & \colhead{Assumed$^a$ (Jy)}
 &	\colhead{Derived$^b$  (Jy)}
}
\startdata
CRL 618		& 450  & 13 $\pm$ 2    &               	\nl
		& 800  & 4.3 $\pm$ 0.2 & 4.3 $\pm$ 0.2 	\nl
		& 1100 & 2.6 $\pm$ 0.1 & 2.8 $\pm$ 0.2 	\nl
		& 1300 & 2.3 $\pm$ 0.1 & 2.7 $\pm$ 0.2 	\nl
		& 2000 & 1.7 $\pm$ 0.1 & 2.2 $\pm$ 0.2 	\nl
						       	\nl
L1551-IRS5	& 450  & 41 $\pm$ 4    &	    	\nl
		& 800  & 5.7 $\pm$ 0.3			\nl
		& 1100 & 2.4 $\pm$ 0.1			\nl
		& 1300 & 1.7 $\pm$ 0.1			\nl
		& 2000 & 0.7 $\pm$ 0.2			\nl
\enddata
\tablerefs
{$^a$Sandell (1994); $^b$run of Sept 1994}  

\end{planotable}
No planets were available for the run of January 1994, so both atmospheric
opacity and flux calibration were estimated 
from observations of CRL 618.  Pointing,
which varied by $<$2-3", was checked at every calibration observation performed
at least once per hour.  Flux densities were also checked using L1551-IRS5.
Assumed flux densities for these two objects were from Sandell (1994),
and are shown in Table 2\footnote{In the September 1994 run we were able to 
check the flux densities of CRL618, and found them to be slightly higher
than the 1991-determined values of Sandell (1994) at longer wavelengths.
(These values are shown in Table 2.)
This might be expected since Sandell noted that the contribution from
free-free emission was slowly increasing at longer wavelengths.  The 
effect of this would be to increase our (1994 January) 
estimated flux densities by 
$\sim$17\% at 1300${\mu}m$ and $\sim$29\% at 2mm. This would
tend to enhance the flux increase we see between the 1992 and
1994 data sets.}.  The quoted errors for these observations (Table 1) include
statistical and calibration uncertainties, the latter of which includes the 
uncertainty in the flux density of CRL 618.

During the run of September 1994, the zenith optical depth at 225 GHz
as measured by the radiometer operated by the nearby
Caltech Submillimeter
Observatory (CSO), was found to be extremely stable, with maximum excursions
ranging from $\tau$ = 0.11 to 0.12.  Thus we were able to point, focus,
and obtain photometry of L1551-IRS5 (only a few degrees away) 
while it was at virtually
the same airmass as GG Tau.  We were also able to observe Mars 
$\sim$2 hours later when it too was at nearly
the same airmass as GG Tau
(1.03 compared to 1.00).  These observations, assuming
flux densities from Sandell (1994) for L1551-IRS5 and from the planetary flux
density model of Wright (1976), 
were then applied {\em directly} to the photometric
observations of GG Tau without having to estimate atmospheric opacity 
or instrumental sensitivity.  (The differences in airmass between Mars
and GG Tau would only introduce an error of $\sim$3\% even assuming 
$\tau_{800{\mu}m} \sim$ 1.)  Consistent values of flux density were found from
both calibrators, and quoted errors reflect both the statistical and
calibration uncertainties.

During the run of December 1994,
we obtained the zenith optical depth at 225 GHz from the CSO radiometer
and found the sky to be stable over the course of 
the observations.
Two observations of Mars at airmasses of 1.5
and 1.1 allowed us to estimate the atmospheric optical depth 
at the observed
wavelengths over the relevant airmass. We also
had observations of L1551 IRS 5 at virtually the same airmass as GG Tau. 
We used these two calibration methods (i.e. deriving $\tau$ and 
sensitivity from
Mars, and direct comparison with L1551-IRS5)
to derive consistent values of flux density for GG Tau.

The data from BSCG were obtained at the 
Institut de Radio Astronomie Millim\'etrique
(IRAM) 30m telescope at Pico Veleta in April 1988,
and were calibrated for flux density using Mars and Uranus.  The effective
wavelength of the observations was 1300${\mu}m$.  BS observed using
a bolometer on the 
CSO in late 1989 and 1990, calibrating for atmospheric opacity using 
HL Tau and for flux density on Mars in 1990 and on W3(OH), CRL618,
IRC+10216, 3C273, 3C84, and HL Tau in 1989 (no planets were available).
The effective wavelengths of these observations were 1056 and 769${\mu}m$
with FWHM of 22'' and 17'' respectively.
The observations from MSWKT were obtained at the JCMT in September 1991 at
effective wavelengths of 790 and 1090 ${\mu}m$ and FWHM of 16.5'' 
and 18.5'' respectively,
calibrating flux density from Mars and atmospheric opacity from CRL618.

\section{Discussion}

\subsection{The Data}

\begin{figure}[t]
\vspace{2.5in}
\centering
   \leavevmode
   \includegraphics{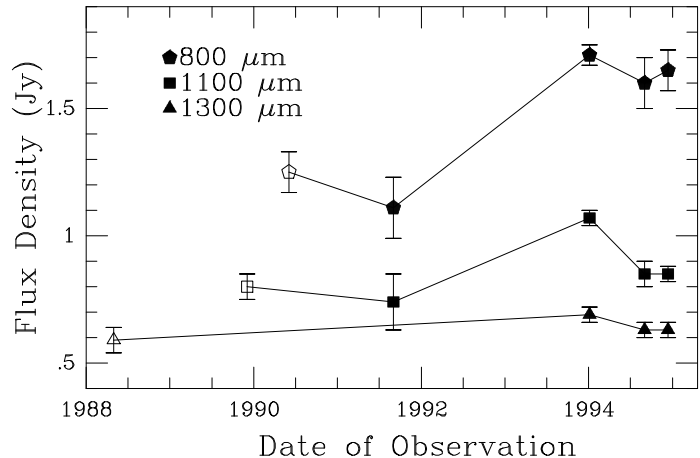}
   \caption{The change of GG Tau's flux at 800, 1100, and 1300 ${\mu}m$.
Solid symbols are from this work and MSWKT, open symbols are from BS or BSCG.}
   \label{ggtau_fig1}
\end{figure}
Figure \ref{ggtau_fig1} 
displays the flux density of GG Tau at 800, 1100 and 1300 ${\mu}m$
as a function of the date of observation.  (The 800${\mu}m$ observation of
BS is an average of observations made in late 1989 and late 1990, and so is 
placed at 1990.4.)
The errors shown on Figure 
\ref{ggtau_fig1} include both statistical and calibration
uncertainties, and are often dominated by the latter.  (Note that these 
errors do not include the uncertainty of the absolute flux of Mars (about 5\%)
based on the model of Wright (1976).)  
It is apparent that from September 1991 to January 1994
the flux density of GG Tau has increased by more than 50\% at 800${\mu}m$
and by $>$40\% at 1100${\mu}m$, but by only $\sim$17\% at 1300${\mu}m$.  From 
January to September 1994, there was a marginal decrease at 800${\mu}m$,
and a decrease to 1991 values at 1100 and 1300 ${\mu}m$.  
It thus appears that the circumbinary disk of GG Tau experienced 
a significant increase in its measured luminosity, 
a ``flare'', between late 1991 and late 1993, affecting predominantly sub-millimeter
wavelengths.  

One must use caution, however, before accepting such a claim.  Errors in
calibration, etc., could account for apparent changes in 
measured flux density.  As noted in {\S}2, all of these data 
except for the run of January 1994 and of December 1989 (from BS) were
calibrated for flux density from Mars, and by using the assumed
Martian flux densities calculated from
the model of Wright (1976).
The December 1989 observation at 1100${\mu}m$ was
calibrated from several sources.  The January 1994 run was calibrated
on CRL618, whose flux density was checked on Mars during the September 1994 run.  
Thus, with minor exceptions, all of the calibration has been 
done in a consistent manner, and we are confident that the derived flux
densities are correct within errors.

\begin{planotable}{cccc}
\tablecaption{Sources Observed 1991 Sept. and 1994 Jan. at 1100${\mu}m$}
\label{sanity-check-tab}
\tablehead{
 \colhead{Source}	&	\colhead{S$_{1100}$ (1991 Sept)}	&
 \colhead{S$_{1100}$ (1994 Jan)}	&	
 \colhead{Ratio S$_{1994}$/S$_{1991}$}	\\
 & \colhead{mJy} & \colhead{mJy} & }
\startdata
04113+2758	& 461 $\pm$ 53  &  510 $\pm$ 40  &  1.11 $\pm$ 0.15 \nl 
04169+2702	& 281 $\pm$ 53  &  230 $\pm$ 30  &  0.82 $\pm$ 0.19 \nl
GG Tau		& 737 $\pm$ 117 & 1070 $\pm$ 30  &  1.45 $\pm$ 0.23 \nl
04361+2547	& 188 $\pm$ 27  & 200 $\pm$  20  &  1.06 $\pm$ 0.19 \nl
04365+2535	& 438 $\pm$ 38  & 390 $\pm$  40  &  0.89 $\pm$ 0.12 \nl
\enddata
\end{planotable}
Five of the sources including GG Tau which were observed 
in September 1991 by MSWKT were 
re-observed at 1100${\mu}m$ during our January 1994 run, providing a 
consistency check.  These data are shown in Table 3.
In four of the five cases, the ratio of 1994 to 1991 flux densities is well
within one standard deviation of being unity.  Only GG Tau has an 
elevated flux density.  Finally we note that the 800${\mu}m$ and 
1100${\mu}m$ flux densities measured by BS and BSCG in 1989/1990 on
different telescopes using different instruments are consistent 
with our 1991 observations and inconsistent with those of post-1991.  
The ratio of our September 1994 800${\mu}m$ data to the
average BS 1989 data  is 1.37$\pm$0.11, and the corresponding 
1100${\mu}m$ data is 1.34$\pm$0.12. 

Other effects could potentially mimic a change in flux density.  
If the source is extended, a smaller aperture 
or beamwidth would detect less of the emission.  However, interferometric
observations, including those of Dutrey, Guilloteau \& Simon (1994), have
resolved the structure of GG Tau, and have found it to be $\lesssim$3",
i.e. much smaller than any of the beamwidths used.
The component to the south associated with GG Tau(c) (the smaller
of the two binary systems) is not visible at 2.7 mm in the Dutrey et al
maps, indicating that there is little contribution to the millimeter
emission from that source.  In addition, the JCMT beam used by MSWKT is
identical to ours, and would cover the same area.

Finally, it must be noted that the observations of BS are made at slightly
{\em shorter} wavelengths (769${\mu}m$ compared to 790${\mu}m$, and
1056${\mu}m$ compared to 1090${\mu}m$).  Since flux density goes as
$S_{\nu} \sim {\lambda}^{-\alpha}$, however, the effect of this should
be to {\em increase} the flux density seen by BS, which is opposite
to the effect seen.

We are thus confident that the increase in flux density of GG Tau seen in
Figure \ref{ggtau_fig1} is {\em real}.

\subsection{Spectral Energy Distribution}

\begin{planotable}{cc}
\tablecaption{Spectral Energy Distributions}
\label{sed-tab}
\tablehead{
 \colhead{Reference}	&	\colhead{$\alpha$}
}
\startdata
BS + BSCG & 1.27 $\pm$ 0.15 \nl
MSWKT     & 1.28 $\pm$ 0.59 \nl
Jan 1994  & 1.73 $\pm$ 0.06 \nl
Sept 1994 & 1.99 $\pm$ 0.10 \nl
Dec 1994  & 1.82 $\pm$ 0.06 \nl
 
\enddata
\tablerefs
{BS = Beckwith \& Sargent (1991);  BSCG = Beckwith et al (1990);  
MSWKT = Moriarty-Schieven et al (1994)}

\end{planotable}
\begin{figure}[t]
\vspace{3.75in}
\centering
  \leavevmode
  \includegraphics{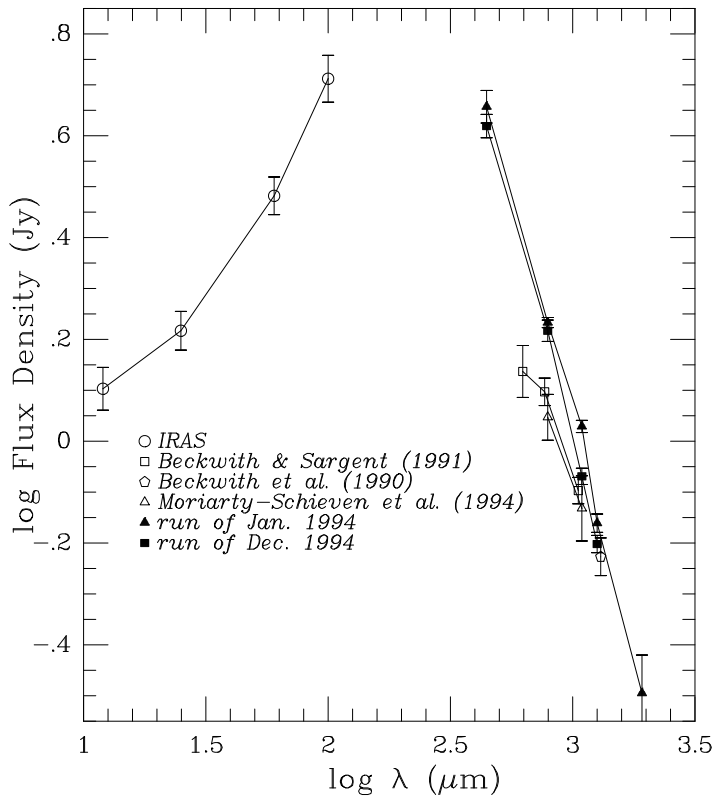}
  \caption{Spectral energy distribution (SED) of GG Tau.  Open symbols 
represent 1983-1991 data, solid symbols represent data obtained in 1994.}
  \label{ggtau_fig2}
\end{figure}
Figure \ref{ggtau_fig2} 
displays the spectral energy distribution (SED) of GG Tau using
the pre-1992 data from Table 1, 
the 1994 January and December data, 
and the IRAS flux densities.  
MSWKT have shown that the slope of the SED at millimeter/sub-millimeter
wavelengths is steeper for embedded sources compared to more evolved,
optically visible, T Tauri stars (like GG Tau).  Indeed GG Tau had the 
shallowest SED of all the sources in their study.  From 
Figure \ref{ggtau_fig2} 
it can be seen that the slope of the SED appears to have
changed as GG Tau has undergone the ``flare''.
In Table 4 we show the slope 
$\alpha$ of the SED (from $S_{\nu} \sim {\lambda}^{-\alpha}$) derived from
each data set (BS includes a 624${\mu}m$ data point plus the data from
BSCG).  The SED appears to have steepened abruptly after the ``flare'', and
continues to be steeper than pre-``flare''.

\subsection{Simple ``Toy'' Source Models}

\begin{figure}[t]
\vspace{2.5in}
\centering
  \leavevmode
  \includegraphics{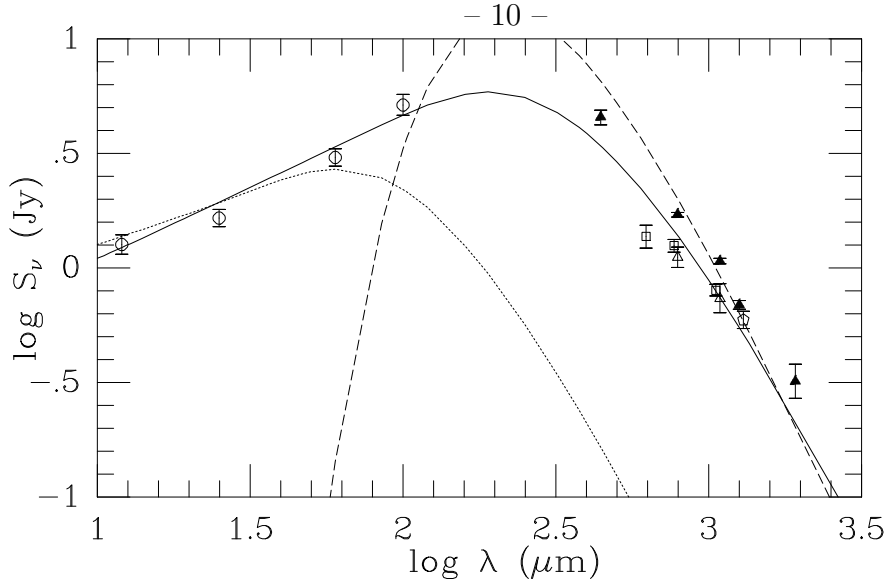}
  \caption{We display the Beckwith et al. (1990) disk model (solid line), 
Dutrey et al disk model (dashed line), and 
a truncated Beckwith et al. disk model (dotted line),
designed to match the IRAS flux out to 50 ${\mu}m$.  The truncated disk is
optically thick at 50 ${\mu}m$ and 15 AU in radius.  The nominal disk
mass however is only 10$^{-3}$ M$_{\sun}$.  Symbols are the same as in 
Figure 2.}
  \label{ggtau_fig3}
\end{figure}
Any model ought to account for a number of effects,
including an increase of the 800 ${\mu}m$ flux density  
of at least 50\% over a timescale
of $\lesssim$2 years, with relatively little effect at longer wavelengths, 
resulting in a 
steepening of the SED during the flare and a more rapid return at 
longer wavelengths to ``normal''.  
Let us consider the existing disk models for GG Tau's continuum emission.
In 1990, Beckwith et al presented a source model for GG Tau
that consisted of a geometrically thin disk, with a large optically
thick regime (out to 45 AU) and an overall disk size of 100 AU.
Their model accounts for the fluxes from 12 to 1300 ${\mu}m$ quite well
(Fig. \ref{ggtau_fig3}, solid line).
However, Dutrey et al (1994) reported that the disk had a large
central 180 AU radius ``hole'', based on 2.6 mm interferometer
observations.   They presented an alternate model, where
a cold dust ring at approximately 180 AU was the dominant emission
source of the 2.6 mm emission. 
Beyond the ring, which contained some 90\% of the disk mass, 
the disk continued with a very flat surface density law.
In addition, they replaced the
optically thin disk with a flared disk, whose scale height
was roughly 30 AU at 180 AU radius.
They argued that the
temperature profile was likely to be T$\propto$r$^{-0.5}$, because
the flared disk would intercept a larger fraction of the stellar
luminosity than spatially thin disk models.  It effectively would act
more like an envelope than a disk when intercepting the stellar
radiation field.  
Artymowicz and Lubow (1995) showed how such
a circumbinary disk might arise and actually be dynamically important
for the evolution of the central binary.

We point out that if the Dutrey et al. idea of a dense dust ring is correct, 
the temperature profile is likely to be 
steeper in the dust ring 
(and the outer disk colder) than their model assumes.  
This occurs
because the dust ring, which has a large extinction,
attenuates the radiation field of the
central stars quite strongly (Wolfire and Churchwell 1994).
While the inner temperature would likely remain at 25 K, the temperature of
the outer disk would fall below the 20 K that is assumed in the Dutrey et al. model.
The outer disk, thus, could be 
effectively shielded from the central heating source.
In addition, their model does not produce any substantial flux
below 100${\mu}m$ (Fig. \ref{ggtau_fig3}, dashed line), 
implying that at least one additional component
(either a smaller circumbinary disk or circumstellar disk) must
be present to explain the IRAS data.  As an illustration of the
type of disk that might be present, we present
in Fig. \ref{ggtau_fig3} (dotted line) 
a disk model which is simply a truncated Beckwith et al. model.
This optically thick (at 50 ${\mu}m$) 
disk has a mass of less than 10$^{-3}$ M$\sun$.
With an outer radius of only 15 AU, it lies
well inside the estimated deprojected separation of 50 AU reported
by Dutrey et al.  Roddier et al. (1996) detected a near-infrared excess
toward both stars in the cavity region, likely due to circumstellar disks.  Also,
the H$\alpha$ emission from GG Tau which led to its classification as a
T Tauri star is believed to be indirect evidence of a circumstellar disk, as
opposed to the circumbinary disk seen at millimeter wavelengths.

\begin{figure}[t]
\vspace{6.5in}
\centering
  \leavevmode
  \includegraphics{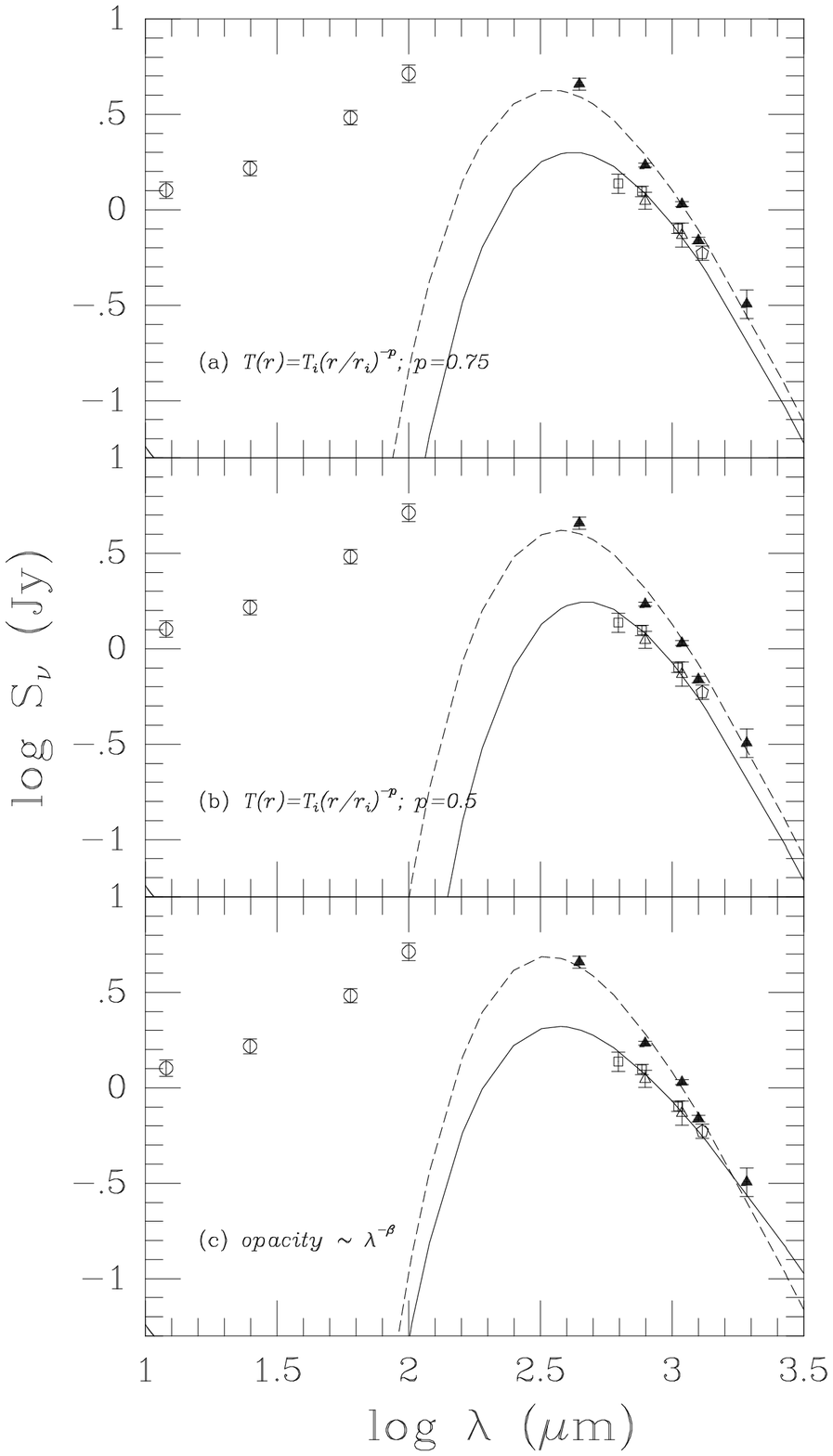}
  \caption{3 ``Toy'' Models: (a) $T(r)=T_i(r/r_i)^{-p}$, $p=0.75$ where
the solid line is a model that fits the pre-brightening phase, and the
dashed line is the same model with the disk luminosity increased by a
factor of 2; (b) $T(r)=T_i(r/r_i)^{-p}$, $p=0.50$, where the solid
line is a model that fits the pre-brightening phase, and the dashed
line is the same model with the disk luminosity increased by a factor
of 2; (c) $opacity \propto \lambda^{-\beta)}$ where $\beta$ is changed from
0.17 in the pre-brightening model (solid line) to 1.0 
in the post-brightening model (dashed line).  The scaling factor
for the opacity law also is changed slightly.  Symbols are the same as in
Figure 2.}
  \label{ggtau_fig4}
\end{figure}
Given the current large uncertainties of the disk(s) physical properties,
we feel that it is unrealistic to construct a detailed model at this
time.  Therefore, we will investigate flat disk models.
For these, we adopt a single temperature power law for the disk,
and the radii indicated by
Dutrey et al., as appropriate for the submillimeter region.  
We adopt the procedures as described in Butner, Natta, \& Evans (1994)
to parameterize the disk properties.
We varied the density, density
profiles, temperature radial profiles, and dust optical properties to get a
sense of how sensitive the predicted spectral energy distributions
are to the input parameters.  
We found the current spectral energy distribution could be fit by 
models within a broad range of parameter space.  
In Fig. \ref{ggtau_fig4}, 
we present three such models, where the ``before'' and
``after'' models are indicated by solid and dashed lines respectively.  
In each case, we limit
ourselves to changing a single input parameter. In Fig \ref{ggtau_fig4}(a),
our model is based on a power law with a 
temperature profile given by T(r)$\propto$ r$^{-p}$; $p = 0.5$.
The only change between the models is the disk's luminosity, or
effectively the disk temperature at the inner radius.  The second 
model (Fig. \ref{ggtau_fig4}(b)) has $p = 0.75$, and again, the only change
between models is an increase in the disk luminosity.
Both fit the pre-brightening phase reasonably well.
By adjusting a single parameter, the total
disk luminosity (or equivalently the temperature at the disk inner
edge), we can then produce the post-brightening observations.
The change in the disk luminosity is modest, going from $\approx$0.01 to
$\approx$0.02 L$_{\sun}$ in both cases. 
The corresponding disk temperatures 
changes are modest, only a few degrees with the disks remaining very cold (10-20 K). 

An alternative to varying the temperature and large-scale disk properties 
is to hold these constant, and instead vary the dust properties.  
In Fig. \ref{ggtau_fig4}(c), 
we present ``pre-'' and ``post-'' brightening models 
where the dust opacity law (parameterized 
as $\propto \lambda ^{-\beta}$) steepens from $\beta =$ 0.2 to $\beta =$ 1.0.
We note that our $\beta =$ 0.2 model, while extremely high, is
in agreement with the estimate of $\beta$ in
Beckwith and Sargent's optically thin disks (their $\beta_p$ models).

It is possible to produce the brightening effect in other ways of course,
such as varying the inner radius of the dust ring.  
The size scales involved (the nominal model 
disk mass is a substantial fraction of 1 M$_{\sun}$
and the radius is order 200 AU), however,  
would argue for longer timescales than
we see for this event.  
We emphasize that the fits in Fig. \ref{ggtau_fig4} are simple ``toy'' 
models intended
as a starting point for detailed calculations, not as a claim
justifying a particular mechanism.
 
\subsection{Possible Outburst Mechanisms}

Our ``toy'' models illustrate that the observations
do not define the parameters sufficiently to identify
physical causes for the observed brightening.  The effect
is dramatic enough, however, that it is interesting to speculate.
What would cause a sudden increase in luminosity (or temperature
across the entire disk) and/or a change in the dust properties 
of a circumstellar disk?  Various possibilities suggest themselves.
For example, an inner disk might still be accreting
material  (Artymowicz and Lubow, 1995).
If this disk were to undergo an outburst, 
it might have increased the energy available
for the outer circumbinary disk to intercept.
To date, no convincing reports of
such dramatic changes have been reported at visual wavelengths
for GG Tau (Herbst, Herbst \& Grossman 1994).  
A standard FU-Ori-type outburst with its 100 times
increase in luminosity certainly has not been seen.  
Another possibility is some sort
of event in the circumbinary disk itself.  
One case that appeals to dynamical events is 
a change in the disk scale height.  
The change in scale height would
permit the disk to intercept more radiation, thus allowing a natural
way to increase the disk's luminosity.
The distances involved  --- a 10 or 15 AU change in scale height is 
required to account for a two-fold increase in total disk luminosity --- 
are large enough to be a concern. 
The required velocities are more than 20 km/sec, and it is
difficult to imagine what type of event would cause parts 
of the disk to achieve such large-scale motions.  
Altering the grain properties would require a sudden
density enhancement for that mechanism to be viable. 
A sudden accretion event in the disk, or interaction with GG Tau/c, the
binary system which lies a few arcseconds to the south (cf Dutrey et al 1994), 
might provide the trigger.
CO(2-1) data presented by Koerner et al. (1993) illustrates that the gas
motions in the area may be more complex than the CO(1-0) indicates,
and the near-IR images of Roddier et al. (1996) show bridge-like 
structures between the outer ring and inner stars.
If there had been an accretion shock as material fell on the
disk, there might be additional energy -- either for heating the
grains or altering their structure enough to affect the optical properties.

Finally, we note that one other T Tauri star, V773 Tauri (HD 283447), is 
known to be variable at mm-wavelengths (Dutrey et al. 1996; Jensen et al. 
(1994); BSCG; Osterloh \& Beckwith 1995).  It is also bright at cm-wavelenghts
(O'Neal et al. 1990; Feigelson et al. 1994) and is rapidly and strongly 
variable at these wavelengths.  Its variability and large brightness
temperature (Phillips, Lonsdale \& Feigelson 1991) 
suggest that the emission is 
synchrotron radiation related to a strong magnetic field.  It is 
conceivable that the sub-millimeter variability of GG Tau may also be
related to non-thermal processes.  However, unlike V773 Tau, GG Tau's
variability seems greatest at short wavelengths, and indeed GG Tau has not
been detected at centimeter-wavelengths despite at least two sensitive
surveys (Bieging, Cohen \& Schwartz 1984; White, Jackson \&
Kundu 1993).  Thus it
seems most likely that the outburst from GG Tau is thermal in nature
and likely related to the circumstellar disk and/or circumbinary ring.

Whatever the explanation proposed, it will require higher resolution
data, particularly at submillimeter wavelengths, to adequately test
the models.

\acknowledgements

While at DRAO, G.M-S. was supported by a research associateship from
the National Research Council of Canada.
H. M. B. gratefully acknowledges the support of the Carnegie
Institute of Washington through a Carnegie Fellowship, 
as well as support from a NASA Origins of 
the Solar Systems program grant (NAGW-4097).  
He further thanks the Max-Planck-Gesellschaft Arbeitsgruppe 
``Staub in Sternentstehungsgebieten'' for their support while
he was on leave at Jena, Germany during the writing of the final
draft of this paper.


\begin{references}

Artymowicz, P., \& Lubow, S. H., 1995, in ``Disks and Outflows Around
Young Stars'', eds. Staude, H. J., Beckwith, S., et al., (Springer Verlag)
in press.

Beckwith, S. V. W., \& Sargent, A. I.  1991, \apj, 381, 250 (BS)

Beckwith, S. V. W., Sargent, A. I., Chini, R. S., \& G\"{u}sten, R.  1990,
\aj, 99, 924 (BSCG)

Beiging, J. H., Cohen, M. \& Schwartz, P. R.  1984, \apj, 282, 699

Butner, H. M., Natta, A., \& Evans, N. J. II 1994, \apj, 420, 326

Dutrey, A., Guilloteau, S., Duvert, G., Prato, L., Simon, M.,
Schuster, K., \& M\'enard, F.  1996, \aap, 309, 493

Dutrey, A., Guilloteau, S., \& Simon, M.  1994, \aap, 286, 149

Elias, J. H.  1978, \apj, 224, 857

Feigelson, E. D., Welty, A. D., Imhoff, C. L., Hall, J. C., Etzel, P. B.,
Phillips, R. B., \& Lonsdale, C. J.  1994, \apj, 432, 373

Hartigan, P., Strom, K. M., \& Strom, S. E.  1994, \apj, 427, 961

Herbig, G., \& Bell, K. R.  1988, {\em Lick Observatory Bull. No. 1111}

Herbst, W., Herbst, D. K., \& Grossman, E. J.  1994, \aj, 108, 1906

Jensen, E. L. N., Mathieu, R. D., \& Fuller, G. A.  1994, \apjl, 429, L29

Leinert, C., Haas, M., Mundt, R., Richichi, A., \& Zinnecker, H.  1991, \aap, 
250, 407

Kawabe, R., Ishiguro, M., Omodaka, T., Kitamura, Y., \& Miyama, S. M.  1993,
\apjlett, 404, L63

Koerner, D. W., Sargent, A. I., \& Beckwith, S. V. W.  1993, \apjlett, 408, L93

Moriarty-Schieven, G. H., Wannier, P. G., Keene, J., \& Tamura, M.  1994, 
\apj, 436, 800 (MSWKT)

O'Neal, D., Feigelson, E. D., Mathieu, R. D., \& Myers, P. C.  1990, \aj, 100,
1610

Osterloh, M., \& Beckwith, S. V. W.  1995, \apj, 439, 288

Phillips, R. B., Lonsdale, C. J., \& Feigelson, E. D.  1991, \apj, 382, 261

Roddier, C., Roddier, F., Northcott, M. J., Graves, J. E., \& Jim, K.
1996, \apj, 463, 326

Sandell, G.  1994, \mnras, 271, 75

Simon, M., \& Guilloteau, S.  1992, \apjlett, 397, L47

Strom, K. M., Strom, S. E., Edwards, S., Cabrit, S., \& Skrutskie, M. F.
1989, \aj, 97, 1451

White, S. M., Jackson, P. D., \& Kundu, M. R.  1993, \aj, 105, 563

Wolfire, M. G., \& Churchwell, E. 1994, \apj, 427, 889

Wright, E. L.  1976, \apj, 210, 250
\end{references}
\end{document}